\documentclass[dvips]{acta}
\usepackage{supertabular,lscape,epsfig}
\usepackage{amssymb}
\usepackage{amsmath}

\SetPages{0}{0}

\SetVol{0}{2014}

\begin{document}

\begin{Titlepage}
\Title{On the GJ 436 planetary system}
\Author{G.~~M~a~c~i~e~j~e~w~s~k~i$^{1}$, ~ A.~~N~i~e~d~z~i~e~l~s~k~i$^{1}$, ~ G.~~N~o~w~a~k$^{2,3}$, ~ E.~~P~a~l~l~\'e$^{2,3}$, ~ B.~~T~i~n~g~l~e~y$^{2,3,4}$, ~ R.~~E~r~r~m~a~n~n$^{5,6}$,  ~ and ~ R.~~N~e~u~h~\"a~u~s~e~r$^{5}$}
{$^1$Centre for Astronomy, Faculty of Physics, Astronomy and Informatics, 
         Nicolaus Copernicus University, Grudziadzka 5, 87-100 Torun, Poland\\ 
         e-mail: gm@astri.umk.pl\\
  $^{2}$Instituto de Astrof\'isica de Canarias, C/ v\'ia L\'actea, s/n, E-38205 La Laguna, Tenerife, Spain\\ 
  $^{3}$Departamento de Astrof\'isica, Universidad de La Laguna, Av. Astrof\'isico Francisco S\'anchez, s/n, E-38206 La Laguna, Tenerife, Spain\\ 
  $^{4}$Stellar Astrophysics Centre, Institut for Fysik og Astronomi, {\AA}arhus Universitet, Ny Munkegade 120, 8000 {\AA}arhus C, Denmark\\ 
  $^5$Astrophysikalisches Institut und Universit\"ats-Sternwarte, 
         Schillerg\"asschen 2--3, 07745 Jena, Germany\\
  $^6$Institute of Applied Physics, Abbe Center of Photonics, Friedrich-Schiller-Universit\"at Jena, Max-Wien-Platz 1, 07743 Jena, Germany} 

\Received{---, 2014}
\end{Titlepage}

\Abstract{The GJ~436 system contains a transiting planet GJ~436~b which is a hot analogue of Neptune on an eccentric orbit. Recently, two additional transiting sub-Earth planets have been postulated in the literature. We observed three transits of GJ~436~b over the course of 3 years using two-meter class telescopes, each with a photometric precision better than one millimagnitude. We studied system dynamics based on the existence of the additional planets. We redetermined system parameters, which were in agreement with those found in the literature.  We refined the orbital period of GJ~436~b and found no evidence of transit timing variations. The orbital motion of the GJ~436~c planet candidate was found to be significantly affected by the planet b with variations in transit times at a level of 20 minutes. As the orbital period of the GJ~436~d planet candidate remains unknown, our numerical experiments rule out orbits in low-order resonances with GJ~436~b. The GJ~436 system with the hot Neptune and additional two Earth-like planets, if confirmed, would be an important laboratory for studies of formation and evolution of planetary systems.}{planetary systems -- stars: individual: GJ~436 -- planets and satellites: individual: GJ~436~b, GJ~436~c, GJ~436~d}


\section{Introduction}

The GJ 436 planetary system, with one confirmed planet and two
proposed planets, may be the first multiple-transiting-planet system
initially discovered from the ground. The host star -- an M3 dwarf
with an age of several Gyr, located 10 pc from the Sun -- was found to
have a Neptune-mass planetary companion on a 2.64 day orbit with the
precise radial velocity (RV) technique (Butler \etal 2004). Additional
Doppler measurements found an orbit that was far from circular, with
an eccentricity of $e_b=0.16\pm0.02$ (Maness \etal 2007) and the
semi-major axis of which is $a_b=0.0285$~ AU. Gillon \etal (2007b)
later detected planetary transits in photometry of the host star.
Combining the spectroscopic and photometric data allowed those authors
to determine the planet's mass of $M_b=22.6 \pm 1.9$ $M_{\oplus}$
(Earth masses) and radius of $R_b=25200 \pm 2200~\rm{km}= 3.95\pm0.34$
$R_{\oplus}$ (Earth radii). These properties showed that GJ~436~b must be
a planet similar to Uranus or Neptune, but much closer to its
host star. Due to this proximity, the planet's
atmosphere is hot, with an equilibrium temperature between 520 and 620
K. These features offer unique opportunities for a number of follow-up
observations conducted with ground-based and space-born facilities.

Maness \etal (2007) and Deming \etal (2007) have postulated that the
non-zero eccentricity could be caused by an unseen planetary
companion. Based on available RV data, Ribas \etal (2008a) found a
sign of additional planet close to the 2:1 mean motion resonance with
GJ~436~b. Coughlin \etal (2008) indicated that orbital inclination,
transit depth, and transit duration may exhibit variations excited by
gravitational influence of another planet in a non-resonant orbit. On
the other hand, dynamical studies of Alonso \etal (2008) and Bean \&
Seifahrt (2008) eliminated the proposed 2:1 configuration, and placed
physical limitations on possible configurations of the second planet
in the system.
  
Using the Spitzer Space Telescope, Stevenson \etal (2012) detected
addtional shallow transit-like features in the light curve of GJ~436. They
propose that these features could be caused by transits of two
additional planets with radii of $\approx 0.7$
$R_{\oplus}$. The orbital period of the planet candidate GJ~436~c
(originally labelled UCF-1.01) was found to be 1.37 d, while the
orbital period of GJ~436~d (UCF-1.02) could not be determined because
only two transits were observed. Assuming a range of bulk densities
typical for terrestrial planets (\ie between 3 and 8~g~cm$^{-3}$), the
masses of both planet candidates were constrained to 0.15--0.40
$M_{\oplus}$ -- much below a detection threshold of current most
advanced RV surveys. The orbital period ratio of GJ~436~b and GJ~436~c
is 1.94, close to a 2:1 orbital resonance which is generally thought
to produce strong transit time variations (TTVs). Recent studies by Lanotte \etal (2014) and  Stevenson \etal (2014) show that the existence of both planet candidates still remains disputable. 

It is believed that future space-based instruments will provide opportunities to confirm the multi-planetary architecture of the GJ~436 system. We note, however, that TTV observations from the ground and dynamical studies may place some constraints on possible planetary configurations. In this paper, we explore the possibility
of the proposed multi-planetary architecture, using new upper limits
on the TTVs of GJ~436~b based on our photometric observations and
combining this with a dynamical model of the system.


\section{Observations and data reduction}

We observed the 2011 Jan 04 transit of GJ~436~b with the 2.2-m
telescope at the Calar Alto Observatory (Spain) as a back-up target
within the program F11-2.2-008, which was focused on transit timing of the WASP-12~b planet. The Calar Alto Faint Object
Spectrograph (CAFOS) in imaging mode was used to acquire the light
curve in the Cousin $R$ filter. To shorten the read-out time, the
original field of view (FoV) was windowed to $5.3' \times
6.5'$. Observations were occasionally affected by passing thin
clouds. We observed a second transit on 2012 Feb 02 with the Nordic Optical
Telescope (NOT) at the Observatorio del Roque de los Muchachos, La
Palma (Spain), as a backup target of the P44-102 observing program
(OPTICON 2011B/003), the goal of which was acquiring high-precision transit light curves for WASP-12~b's transits. The Andalucia Faint Object Spectrograph and
Camera (ALFOSC) was used in an imaging mode, windowed to the
$6.4' \times 6.4'$ FoV. Photometric time series was taken in the
Bessel $R$ filter under excellent weather conditions.

Both instruments used $2\times2$ binning for faster
readout. Autoguiding guaranteed that stellar centroids did not change
their locations on the detector matrix during each run. We defocused
the telescopes significantly, creating doughnut-like stellar profiles
and spreading the light over many pixels, which enabled longer
exposure times and reduced the impact of flat-fielding
imperfections. This also improves duty cycle, as the ratio of time
during exposure versus read-out is improved, allowing more photons to
be gathered during transit and thereby increasing the photometric
efficiency. We reduced the observations with standard procedures,
including de-biasing and flat-fielding using sky flats. We performed
differential aperture photometry with respect to nearby
comparison stars BD+27~2046 and TYC~1984-1884-1. Radii of the aperture
and background ring were empirically optimized to produce the smallest
scatter in the out-of-transit light curve.

We obtained an additional transit light curve from observations made
on 2014 Mar 23 from low-resolution spectra acquired with the
NOT/ALFOSC. For these observations we chose the $2 \times 2$ binning
mode, a readout speed of 200 pixel/sec with a gain of 0.327
e$^{-}$/ADU and a readout noise of 4.2 e$^{-}$/pixel. We used ALFOSC
in its long-slit spectroscopic mode, selecting the grism \#4 which
covers the spectral range 3200 -- 9100 {\AA} and a custom-built slit
of a width of 40 arcsec. We chose BD+27~2046 as reference, which has
similar brightness and is located at a distance of 3.81 arcmin from
GJ~436. The position angle of the reference star with respect to the
target was equal to $37.39^{\circ}$. Observations began at 01:31 UT
(35 minutes before ingress) and ended at 03:18 UT (10 minutes after
egress). The exposure time was set to 60 seconds and the readout time
of instrument was 9 s, meaning we collected approximately one spectrum
every 69 s. The data reduction (bias and flat-field corrections,
extraction of the spectra and corresponding calibration arcs as well
as wavelength calibration using the He and Ne lamps) was made using an
IRAF \footnote{IRAF is distributed by the National Optical Astronomy
Observatory, which is operated by the Association of Universities for
Research in Astronomy (AURA) under cooperative agreement with the
National Science Foundation.} script written for NOT/ALFOSC long-slit
data. Optimal extraction of the spectra was obtained using an aperture
width of $\pm$10 binned pixels, which corresponds to 7.6 arcsec on the
detector. This is 2 to 5.5 times the raw seeing during the
observations (1.4 -- 3.7 arcsec). The light curve was constructed
using 1/3 of spectra of the target and reference star centered at the
maximum of the GJ~436 spectrum. The maximum of the spectral energy
distribution was found at 790 nm that corresponds to the central
passband of the photometric $I$ filter.

Differential atmospheric extinction and differences in spectral types
of the target and comparison stars, as well as instrumental effects
caused by the field derotator in NOT data, are expected to produce
photometric trends, whose time scale is similar to the duration of the
run. The de-trending procedure was done with the {\sc jktebop} code
(Southworth \etal 2004a,b) by fitting a second-order polynomial
function of time along with a trial transit model and then
subtracting the resulting polynomial from the light curve. Magnitudes
were transformed into fluxes, which were normalized to unity outside
of the transit. The timestamps in geocentric Julian dates in
coordinated universal time (UTC) were provided by a GPS system and
verified with the network time protocol software, then
converted to barycentric Julian dates in barycentric dynamical time
(BJD$_{\rm{TDB}}$, Eastman \etal 2010). The observing details and data quality characteristics are given in Table~1.

\MakeTable{c l l c c c c}{12.5cm}{New transit light curves reported for GJ~436~b: date UT is given for the middle of the transit, $X$ shows a course of changes in airmass during transit observations, $t_{exp}$ is the exposure time, $\Gamma$ is the median number of exposures per minute, $pnr$ is the photometric scatter in millimagnitudes per minute of observation, as defined by Fulton \etal (2011).}
{\hline
\# & Date UT  &  Telescope & $X$  & $t_{exp}$ (s) & $\Gamma$ & $pnr$ (mmag)\\ 
\hline
   1 & 2011 Jan 04 & 2.2-m Calar Alto & $1.08\rightarrow1.02\rightarrow1.07$  & $12$ & 1.76  &  0.6\\ 
   2 & 2012 Feb 02 & 2.6-m NOT & $1.08\rightarrow1.00\rightarrow1.06$  & $20$ & 2.07  &  0.3\\    
   3 & 2014 Mar 23 & 2.6-m NOT & $1.01\rightarrow1.18$  & $60$ & 0.87  &  0.7\\ 
\hline
}


\section{Results}


We used different codes to model the transit light curves and analyze
the RV and timing datasets. We obtained the final results of this
study by iterating the different codes until we reached
convergence. Sections 3.1 and 3.2 contain a description of the steps
involved in each iteration.

\subsection{Transit light curves}

We modeled the new sub-millimagnitude precision light curves
simultaneously with the Transit Analysis
Package\footnote{http://ifa.hawaii.edu/users/zgazak/IfA/TAP.html}
({\sc TAP}, Gazak \etal 2012). This code employs the Markov Chain
Monte Carlo (MCMC) method, including the Metropolis-Hastings algorithm
and a Gibbs sampler, to find the best-fit parameters of a transit
light curve approximated by the analytical model of Mandel \& Agol
(2002). The time-correlated noise in data (so called red noise) is
investigated with the Carter \& Winn (2009) wavelet
parametrization. This approach is known to provide conservative
uncertainty estimates. TAP parametrizes the flux distribution across
the stellar disk with a quadratic limb darkening (LD) law (Kopal
1950). The values of linear and quadratic LD coefficients, $u_1$ and
$u_2$ respectively, were linearly interpolated from tables of
Claret \& Bloemen (2011) with an on-line
tool\footnote{http://astroutils.astronomy.ohio-state.edu/exofast/limbdark.shtml}
of the \textsc{EXOFAST} package (Eastman \etal 2013). The stellar
parameters for GJ~436 were taken from von Braun \etal (2012), assuming
solar metallicity.

The fitting procedure kept the orbital inclination $i_{b}$, the
semimajor-axis scaled by stellar radius $a_{b}/R_{*}$, and the
planetary to stellar radii ratio $R_{b}/R_{*}$ as free parameters,
linked together for all light curves.  The mid-transit times were
determined independently for each light curve. The orbital period was
fixed at a value from a refined ephemeris. The LD coefficients were
allowed to vary around the theoretical values under the Gaussian
penalty of $\sigma = 0.05$, independently for all three datasets. This
allowed us to account not only for differences between $R$ and $I$
bands, but also for any possible differences in instrumental
implementation of $R$-band filters. The orbital eccentricity $e_b$ and
longitude of periastron $\omega_b$ were taken from the dynamical model
discussed in Sect.~3.2. The new light curves with the best-fit transit
model are plotted in Fig.~1.

Our new transit observations prolong the timespan covered by
observations and allow the transit ephemeris to be refined. The
derived mid-transit times were combined with the literature ones to
calculate new reference epoch $T_0$ and orbital period $P_b$.

The parameters of our best-fit model are given in Table~2, together
with recent literature results for comparison. We also list the resulting
system properties: the transit parameter $b_{{b}}$, defined as
\begin{equation}
     b_{{b}}=\frac{a_b}{R_{*}}\frac{1-e_b^2}{1+e_b\cos\omega_b}\cos{i_{b}}\, , \; 
\end{equation}
and the mean stellar density $\rho_{*}$, which can be directly calculated
from transit observable properties with a formula derived from Kepler's
third law
\begin{equation}
     \rho_{*} = \frac{3\pi}{G P_{b}^2} \left(\frac{a_{{b}}}{R_{*}}\right)^3\, , \; 
\end{equation}
where $G$ is the gravitational constant. 

\MakeTable{lccc}{12.5cm}{Parameters of the GJ~436 system derived from modeling transit light curves.}
{\hline
Parameter &  This work &  von Braun \etal (2012)  &  Lanotte \etal (2014) \\
\hline
$i_{b}$ ($^{\circ}$) & $86.44^{+0.17}_{-0.16}$ &  $86.6^{+0.1}_{-0.1}$ &  $86.858^{+0.049}_{-0.052}$ \\
$a_{b}/R_{*}$ & $13.73^{+0.46}_{-0.43}$ &  -- & $14.54^{+0.14}_{-0.15}$\\
$R_{b}/R_{*}$ & $0.0822^{+0.0010}_{-0.0011}$ &  $0.0833\pm0.0002$ & -- \\
$b_{b}$ & $0.736^{+0.042}_{-0.041}$ &  $0.853^{+0.003}_{-0.003}$ &  $0.7972^{+0.0053}_{-0.0055}$ \\
$\rho_{*}$  $(\rho_{\odot})$ & $4.97^{+0.50}_{-0.47}$ &  $5.37^{+0.30}_{-0.27}$  &  $5.91^{+0.17}_{-0.18}$ \\
$T_{0}$ (BJD$_{\rm{TDB}}$) & $2454510.80162\pm0.00007$ &  $2454510.80096\pm0.00005$ & --\\
$P_{b}$ (d) & $2.64389754\pm0.00000043$ &  $2.64389826^{+0.00000056}_{-0.00000058}$ &  $2.64389803^{+0.00000027}_{-0.00000025}$ \\
\hline
}

\begin{figure}[t]
\begin{center}
\includegraphics[width=1.0\textwidth]{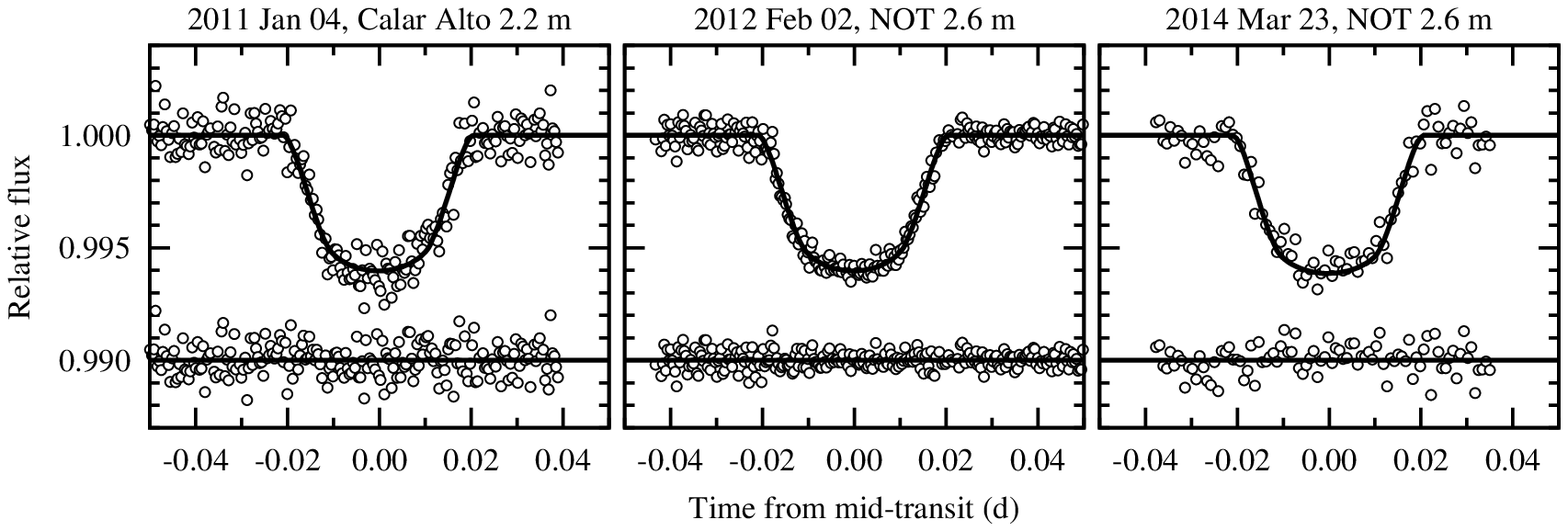}
\end{center}
\FigCap{New transit light curves for GJ~436~b with the best-fit model, plotted with solid lines. The residuals are shown in bottom parts of the panels.}
\end{figure}


\subsection{Dynamical model with the GJ~436~c planet candidate}

The linear fit to mid-transit times for GJ~436~b results in a reduced
$\chi^2$ equal to 5.5, which indicates a marked departure from a
linear ephemeris (Stevenson \etal 2012). The periodogram analysis
reveals no statistically significant periodic signal, so stellar
activity, systematic effects, or underestimated timing errors may be a
source of a spurious timing variations. An upper limit on the
amplitude of any periodic signal in transit timing was found to be
0.0002 d.

Mid-transit times for the GJ~436~c planet candidate show noticeable
departure from a linear ephemeris (Stevenson \etal 2012) that suggests
planet's orbital motion is perturbed by the planet b. To study the
mutual interactions between both planets, a two-planet dynamical model
was constructed with the Systemic code in version 2.16
(Meschiari \etal 2009). We used 113 RV measurements from Knutson \etal
(2014), acquired with the HIRES echelle spectrometer at the Keck I
telescope between January 2000 and December 2010. We also used 171
HARPS RVs obtained between January 2009 and April 2010 from
Lanotte \etal (2014). For GJ~436~b, we used mid-transit times for 37
epochs available in the literature (Gillon \etal 2007a,b;
Shporer \etal 2009; C\'aceres \etal 2009; Deming \etal 2007;
Alonso \etal 2008; Knutson \etal 2011; Pont \etal 2009; Bean \etal
2008; Ribas \etal 2008b; Coughlin \etal 2008; Ballard \etal 2010;
Beaulieu \etal 2011) and 3 new determinations reported in this
paper. In addition, we used 15 mid-occultation times from
Stevenson \etal (2010) and Knutson \etal (2011). The best-fit
Newtonian solution was found with a differential evolution algorithm
using 5000 steps, followed by a number of iterations of
Levenberg-Marquardt optimization. The Runge-Kutta-Fehlberg (RK45)
algorithm was used to integrate equations of motion with an accuracy
requirement of $10^{-16}$. The bootstrap method with $10^3$ trials was
used to estimate parameter uncertainties, calculated as median
absolute deviations. The orbital periods, eccentricities, and
arguments of periapsis were left free for both planets. The mass of the
planet candidate was fixed at the value of 0.28 $M_{\oplus}$ (Earth
masses) taken from Stevenson \etal (2012). The mass of planet b was
allowed to vary. Orbital inclinations were taken from our transit
light-curve analysis (Sect.~3.1) for GJ~436~b and from Stevenson \etal
(2012) for the planet candidate. The parameters of the best-fit
dynamical model are listed in Table 3.

\MakeTable{lcc}{12.5cm}{Orbital parameters for the GJ~436 b planet
and GJ~436 c planet candidate from the two-planet dynamical model.
The values are given for the epoch JD 2455959.}
{\hline
Parameter &  GJ~436 b &  GJ~436 c \\
\hline
Orbital period (d) & $2.64388312\pm0.00000057$ &  $1.365960\pm0.000012$ \\
Semi-major axis (AU) & $0.0291\pm0.0015$ & $0.01871^{+0.00097}_{-0.00094}$  \\
Orbital eccentricity & $0.13827\pm0.00018$ &  $0.1166\pm0.0046$ \\
Longitude of periastron (deg) & $351.00\pm0.03$ &  $36.50\pm0.41$ \\
RV amplitude  (m/s) & $17.09\pm0.22$ &  $0.3^{*}$ \\
Mass ($M_{\oplus}$) & $22.1\pm2.3$ &  $0.28^{**}$ \\
\hline
\multicolumn{3}{l}{$^{*}$ predicted value}\\
\multicolumn{3}{l}{$^{**}$ value taken from Stevenson \etal (2012).}\\
}

The SWIFT's Regularized Mixed Variable Symplectic (RMVS) integrator
was used to trace the evolution of orbital parameters, which exhibit
oscillatory patterns as a function of time. The best-fit dynamical
model is stable in a timescale of $10^6$ yr (over 130 million orbits
of GJ~436~b) and yields strong constraints on the eccentricity of the
GJ~436~c planet candidate. Its best-fit value is similar to that one
for GJ~436~b but it is expected to oscillate between a value as small
as 0.02 and 0.19 with a period of 35.4 yr (Fig.~2a), at anti-phase
with marginal variations in $e_b$ (between 0.137 and 0.139; the range
is smaller by a factor of the mass ratio that is a consequence of
conservation of momentum). The predicted variation in orbital
inclination for GJ~436~c is $2.6^{\circ}$ (Fig.~2b). The inclination
of GJ~436~b's orbit was found to decrease with a rate of 0.025 degree
per century -- far under the detection threshold of current transit
observations.  The ratio between the orbital periods of both planets,
$P_b/P_c=1.94$, suggests the planets are close to a 2:1
commensurability and could be trapped in a mean motion resonance. The
evolution of the difference between arguments of periastron, defined
as $\Delta\omega=\omega_b-\omega_c$, is plotted in Fig.~2c. The
periastrons were found to be in apsidal alignment around $0^{\circ}$
and librate with an amplitude of $56^{\circ}$. Such conditions are
generated by the linear secular coupling and prevent both planets from
close encounters which could destabilize the system. The
eccentricity-type resonant angles, defined as a linear combination of
mean longitudes $\lambda$ and arguments of periastron,
\begin{align}
 \theta_{{b}} & = \lambda_c - 2\lambda_b+\omega_b \\
 \theta_{{c}} & = \lambda_c - 2\lambda_b+\omega_c
\end{align}
show no libration, so both planets are not in a dynamical resonance.      

\begin{figure}[t]
\begin{center}
\includegraphics[width=0.7\textwidth]{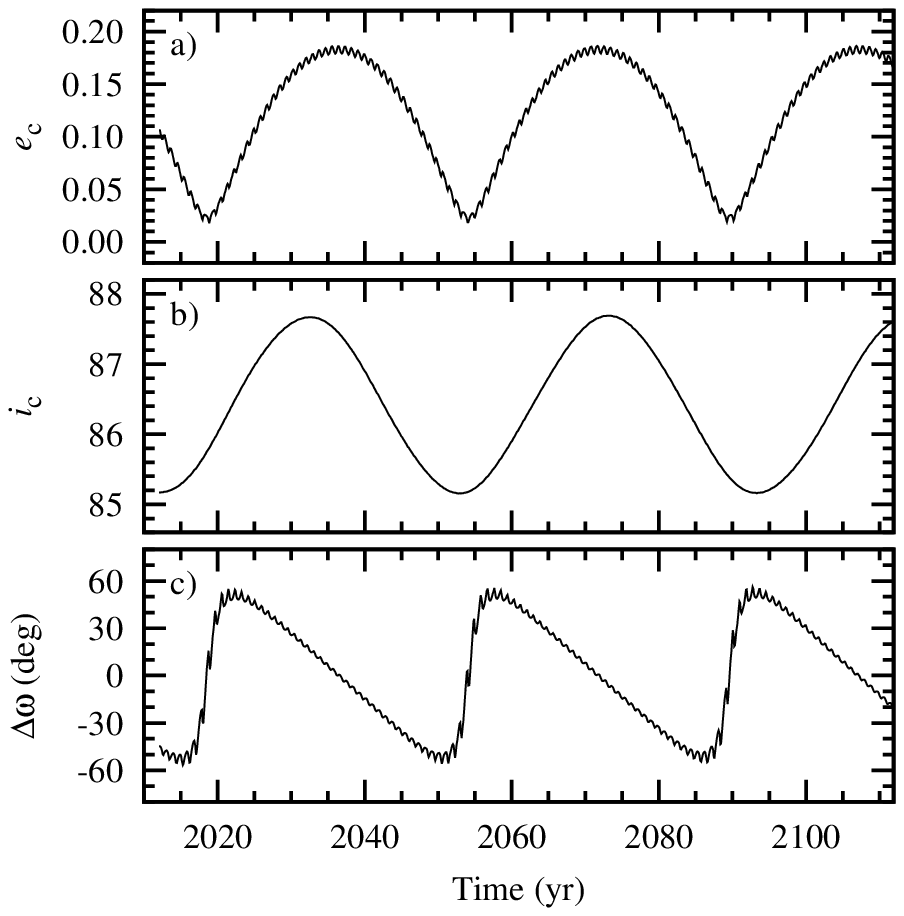}
\end{center}
\FigCap{Evolution of the orbital eccentricity and inclination for the
GJ~436~c planet candidate (panels a and b) and difference in arguments 
of periapsis between GJ~436~b and c.}
\end{figure}

Figure~3 shows transit timing residuals for GJ~436 b and the GJ~436 c
planet candidate, produced by mutual gravitational interactions. For
both planets, the signal is periodic with a period of 41.2 d. The
range of variation is substantial for the GJ~436 c planet candidate
with a value of almost 20 min. Timing of GJ~436 b is predicted to be
modulated with an amplitude of 16 s -- a signal which is under the
detection threshold of the current timing dataset.

\begin{figure}[t]
\begin{center}
\includegraphics[width=0.9\textwidth]{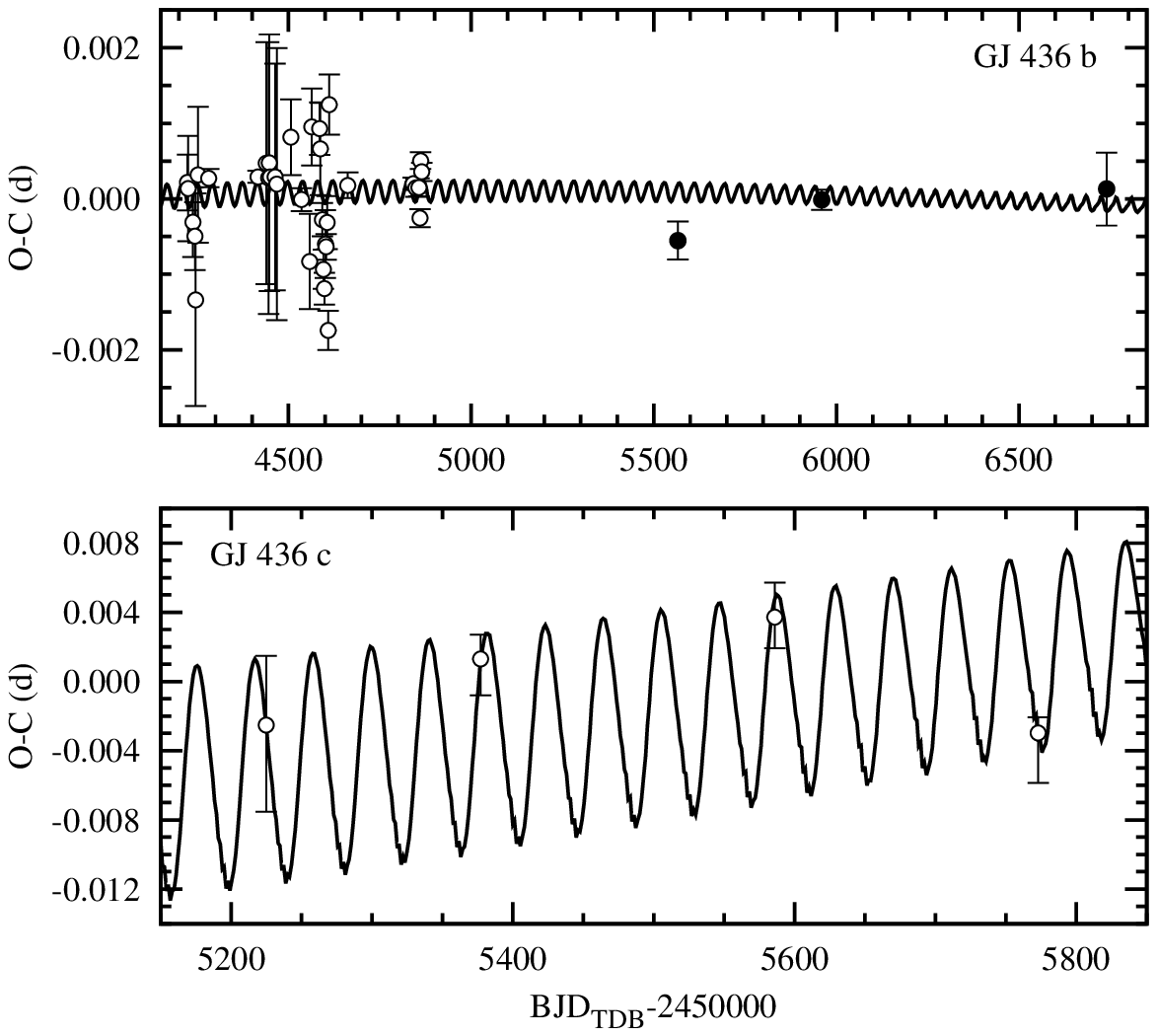}
\end{center}
\FigCap{Timing residuals for transits of the GJ~436 b planet and the
GJ~436 c planet candidate, resulting from mutual gravitational interactions.}
\end{figure}

\subsection{Constraints on the orbital period of the GJ~436 d planet candidate}

The period of time between the two observations of the transits of
candidate GJ~436~d, $\tau_d=151.5$ d, must be a multiple of a true
orbital period $P_d$. Stevenson \etal (2012) estimate the upper limit
for $P_d$ to be 5.56 d, based on 1-$\sigma$ uncertainty in transit
duration of the planet candidate. We note however, that the fact that
the orbital inclination is unknown weakens their argument. Periods
shorter than $P_b$ may be excluded because no transit signature was
found in the 2.4-day long continuous light curve of GJ~436 acquired with the
Spitzer Space Telescope at 8 $\mu$m (Stevenson \etal 2012). We used
GJ~436~b's transit timing and system's dynamical stability to put
constraints on $P_d$.

Using the Systemic code, we conducted a numerical experiment in which
the planet~d was inserted into the two-planet system derived in
Sect.~3.2. The mass of the planet d was set to 0.27 $M_{\oplus}$ as
given by Stevenson \etal (2012). The orbital eccentricity, $e_d$,
varied between 0.0 and 0.3 with a step of 0.05. The initial value of
the argument of periastron was set equal to ${\omega}_b$ and initial
orbital longitude was shifted by $180^{\circ}$ with respect to the
value for GJ~436~b at the epoch 0. The orbital inclination of the
planet d was the same as for GJ~436~b. The value of $P_d$ satisfied
the relation
\begin{equation}
     \tau_d=n \cdot  P_d\, , \; 
\end{equation}
where $n$ is 1, 2, 3... Each configuration was integrated with the
RK45 algorithm, covering 2500 days, \ie the time span of the transit
observations for GJ~436~b. The times of mid-transit times were
extracted and the amplitude of periodic deviations from a linear
ephemeris was calculated. In addition, each configuration was checked
for the dynamical stability in $10^3$ yr with the SWIFT's RMVS
integrator. Configurations which were found to be unstable, mainly
with $P_d<3/2P_b$, were skipped.

The exemplary results for $e_d$ equal to 0.05, 0.15, and 0.25 are
shown in Fig.~4. Configurations with $P_d/P_b$ close to 3:2 and 2:1
resonances would generate TTV signals above a detection
threshold even for low-eccentricity orbits of the planet d. For
greater eccentricities, configurations with resonances close to
3:1 or 5:2 would produce detectable TTVs.

\begin{figure}[t]
\begin{center}
\includegraphics[width=0.8\textwidth]{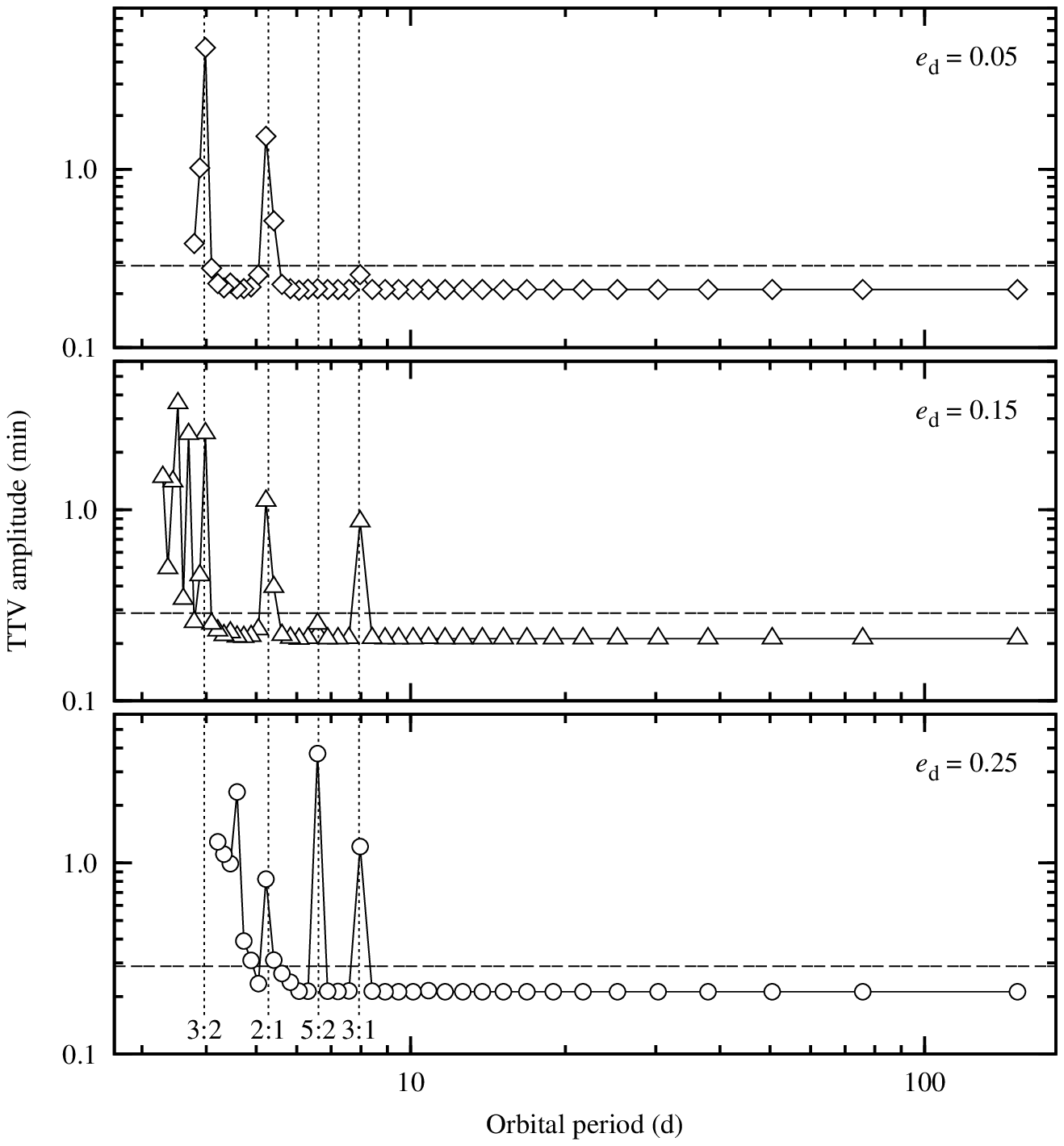}
\end{center}
\FigCap{Amplitude of a hypothetical TTV signal for GJ~436~b induced by the GJ~436~d planet candidate as a function of possible values of $P_d$ for selected values of $e_d$. Only configurations that are stable for a time scale of $10^3$ yr are shown. The horizontal dashed lines show a detection threshold based on observations. Vertical dotted lines mark selected $P_d$ to $P_b$ resonances.}
\end{figure}


\section{Concluding discussion}

Our new observations of GJ~436~b's transits allowed us to refine
transit ephemeris and to redetermine system parameters. They prolong
the timespan covered by observations by a factor of 3. They were
combined with photometric and Doppler data from the literature in
order to study dynamics of the system with two low-mass planet
candidates.

The interferometric measurements for the GJ~436 host star give a
precise value of stellar radius of $R_*=0.455\pm0.018$~$R_{\odot}$ (von Braun \etal 2012),
and when combined with a mean stellar density of
$\rho_{*}=4.97^{+0.50}_{-0.47}$~$\rho_{\odot}$ results in a stellar mass of
$M_{*}=0.47\pm0.07$~$M_{\odot}$. This value is between
 $0.452^{+0.014}_{-0.012}$~$M_{\odot}$ and $0.507^{+0.071}_{-0.062}$~$M_{\odot}$ reported by Torres (2007) and von Braun \etal (2012), respectively. It is also still consistent within a 1-$\sigma$ range with the value of $0.556^{+0.071}_{-0.065}$~$M_{\odot}$ given by Lanotte \etal (2014). The deduced stellar surface
gravity is $\log g_{*}=4.792^{+0.047}_{-0.044}$ in cgs units,
consistent with $\log g_{*}=4.83\pm{0.03}$ derived by von Braun \etal
(2012). Our determinations of the GJ~436~b's radius
$R_b=0.372\pm0.015$~$R_{Jup}$ and mean density
$\rho_{b}=1.35\pm0.22$~$\rho_{Jup}$  agree with the recent literature
values within 1-$\sigma$ ($0.369\pm0.015$~$R_{Jup}$ and $1.55^{+0.12}_{-0.10}$~$\rho_{Jup}$ reported by von Braun \etal 2012, and $0.366\pm0.014$~$R_{Jup}$ and $1.6$~$\rho_{Jup}$ given by
Lanotte \etal 2014). Our value of planetary mass $M_b=22.1\pm2.3$~$M_{\oplus}$ is in a perfect agreement with $M_b=22.6 \pm 1.9$ $M_{\oplus}$ given by Gillon \etal (2007b), but seems to be slightly underestimated comparing to $24.8^{+2.2}_{-2.5}$~$M_{\oplus}$ of von Braun \etal (2012) and $25.4^{+2.1}_{-2.0}$~$M_{\oplus}$ of Lanotte \etal (2014). This is a direct consequence of higher stellar mass determined in both studies.

Mid-transit times reported in this study follow a linear ephemeris
with residuals smaller than $2\sigma$. The transit from 2011 Jan 04
(JD 2455565.7), the most outlying from the linear ephemeris in our
sample, could be affected by stellar activity at the ingress phase
because photometric residuals from the transit model seem to exhibit
some distortion. The lack of a periodic or semi-periodic TTV signal
indicates that GJ~436~b is not noticeably perturbed by gravitational
interactions with other bodies in the system. On the other hand, the
orbital motion of the GJ~436~c planet candidate is significantly
influenced by GJ~436~b. Our numerical model predicts that periastrons
of both planets are in apsidal alignment. The TTV signal for GJ~436~c
is expected to be at the level of 20 min and constrains the orbital
eccentricity of the planet. Further precise transit observations for
GJ~436~c will shed new light for system dynamics.

The orbital period of the GJ~436~d planet candidate remains unknown
but the lack of the TTV signal for GJ~436~b put some constraints on
it. Configurations in which both bodies are close to low-order
resonances are unlikely because they would generate detectable TTVs
for GJ~436~b. A system with planets out of resonance is in line with
statistical studies based on data from the space-based {\it Kepler}
transit survey, which show that configurations with planets in close
proximity to a resonance are not favoured (Fabrycky \etal 2014).


\Acknow{We would like to thank NOT and Calar Alto staff for their
help during observing runs. GM acknowledges the financial support from the Polish Ministry of Science and Higher Education through the Iuventus Plus grant IP2011 031971. GM and AN acknowledge funding from the European Community's Seventh Framework Programme (FP7/2007-2013) under grant agreement number RG226604 (OPTICON).
AN is supported by the Polish Ministry of Science and Higher Education grant N N203 510938.
We would like to thank the German national science foundation Deutsche Forschungsgemeinschaft
(DFG) for financial support for the Calar Alto run in program NE 515 / 44-1. RE would like to thank DFG for support in the Priority Programme SPP 1385 in project NE 515 / 34-1 and the Abbe School of photonics for the grant.
The research is based on data collected with the Nordic Optical Telescope, operated on the island of La Palma jointly by Denmark, Finland, Iceland, Norway, and Sweden, in the Spanish Observatorio del Roque de los Muchachos of the Instituto de Astrof\'{\i}sica de Canarias. Some observations were made at the Centro Astron\'omico Hispano Alem\'an (CAHA), operated jointly by the Max-Planck Institut f\"ur Astronomie and the Instituto de Astrof\'{\i}sica de Andaluc\'{\i}a (CSIC).
Data presented here were obtained with ALFOSC, which is provided by the Instituto de Astrof\'{\i}sica de Andaluc\'{\i}a (IAA) under a joint agreement with the University of Copenhagen and NOTSA.}

\end{document}